\documentclass[proceedings]{JHEP37}
\usepackage{eurosym}
\usepackage{amssymb}
\usepackage{amsfonts}
\usepackage{amsmath}
\usepackage{epsfig}

\newbox\mybox

\newcommand\fverb{\setbox\mybox=\hbox\bgroup\verb}
\newcommand\fverbdo{\egroup\medskip\noindent\fbox{\unhbox\mybox}\ }
\newcommand\fverbit{\egroup\item[\fbox{\unhbox\mybox}]}
\conference{A unifying E2-quasi-exactly solvable model}
\abstract{A new non-Hermitian E2-quasi-exactly solvable model is constructed containing two previously known models of this type as limits in one of its three parameters. We identify the optimal finite approximation to the double scaling limit to the complex Mathieu Hamiltonian. A detailed analysis of the vicinity of the exceptional points in the parameter space is provided by discussing the branch cut structures responsible for the chirality when exceptional points are surrounded and the structure of the corresponding energy eigenvalue loops stretching over several Riemann sheets. We compute the Stieltjes measure and momentum functionals for the coefficient functions that are univariate weakly orthogonal polynomials in the energy obeying three-term recurrence relations.}

\title{A unifying E2-quasi exactly solvable model}
\author{Andreas Fring \\
Department of Mathematics, City University London,\\
Northampton Square, London EC1V 0HB, UK\\
E-mail: a.fring@city.ac.uk}

\input{tcilatex}
\begin{document}

\section{Introduction}

In addition to the interesting mathematical aspect of enlarging the set of $%
sl_{2}(\mathbb{C})$ \cite{Turbiner00,Tur0} to $E_{2}$-quasi-exactly solvable
models \cite{E2Fring}, the latter type also constitutes the natural
framework for various physical applications in optics where the formal
analogy between the Helmholtz equation and the Schr\"{o}dinger equation is
exploited \cite%
{Muss,MatMakris,Guo,OPMidya,MatHugh,MatHughEva,DFM,DFM2,MatLongo,constint}.
Furthermore, a special case of these systems with a specific representation
corresponds to the complex Mathieu equation that finds an interesting
application in nonequilibrium statistical mechanics, where it corresponds to
the eigenvalue equation for the collision operator in a two-dimensional
classical Lorentz gas \cite{KazukiTalk,thesisZhang}.

Here we are mainly concerned with the extension of quasi-exactly solvable
models \cite{KhareM,BijanMQR,BenMou,BijanQR,E2Fring} to non-Hermitian
quantum mechanical systems \cite{Urubu,Bender:1998ke,Benderrev,Alirev}
within the above mentioned scheme. So far two different types of $E_{2}$%
-models have been constructed in \cite{E2Fring,E2Fring2} and the main
purpose of this manuscript is to investigate whether it is possible to
construct a more general model that unifies the two. We show that this is
indeed possible by combining the two models and introducing a new parameter
into the system that interpolates between the two. In a similar fashion as
the previously constructed models, also this one reduces in the double
scaling limit to the complex Mathieu equation. As that equation is not fully
explored analytically this limit provides an important option to obtain
interesting information about the complex Mathieu system. On the other hand,
for some applications it may also be sufficient to study an approximate
behaviour for some finite values of the coupling constants. For that purpose
we identify the parameter for which the general model is the optimal
approximation for the complex Mathieu system.

Our manuscript is organized as follows: In section 2 we introduce the
general unifying model involving three parameters. We determine the
eigenfunctions by solving the standard three-term recurrence relations for
the coefficient functions and determine the energy eigenfunction from the
requirement that the three-term recurrence relations reduce to a two-term
relation. We devote section three to the study of the exceptional points and
their vicinities in the parameter space. The explicit branch cut structure
is provided that explains the so-called energy eigenvalue loops. In section
4 we compute the central properties of the weakly orthogonal polynomials
entering as coefficient functions in the Ansatz for the eigenfunctions, i.e.
their norms, the corresponding Stieltjes measure and the momentum
functionals. We state our conclusions in section 5.

\section{A unifying E2-quasi-exactly solvable model}

The general notion \cite{Turbiner00,Tur0} underlying solvable Hamiltonian
systems is that its Hamiltonian operators $\mathcal{H}$ acting on some
graded space $V_{n}$ as $\mathcal{H}:V_{n}\mapsto V_{n}$ preserves the flag
structure $V_{0}\subset V_{1}\subset V_{2}\subset \ldots \subset
V_{n}\subset \ldots $ A distinction is usually made between exactly and
quasi-exactly solvable, depending on whether the structure preservation
holds for an infinite or a finite flag, respectively. Here we are concerned
with the latter. Lie algebraic versions of Hamiltonians in this context are
usually taken to be of $sl_{2}(\mathbb{C})$-type \cite{Turbiner00,Tur0}, but
as recently proposed \cite{E2Fring,E2Fring2}, they may also be taken to be
of a Euclidean Lie algebraic type, thus giving rise to qualitatively new
structures.

At present two different types of $E_{2}$-quasi-exactly solvable models were
identified 
\begin{eqnarray}
\mathcal{H}_{E_{2}}^{(1)} &=&J^{2}+\zeta ^{2}(u^{2}-v^{2})^{2}+2i\zeta
N(u^{2}-v^{2}),\qquad \qquad \zeta ,N\in \mathbb{R},\quad  \label{H1} \\
\mathcal{H}_{E_{2}}^{(0)} &=&J^{2}+\zeta uvJ+2i\zeta N(u^{2}-v^{2}),\quad
\label{H0}
\end{eqnarray}%
in \cite{E2Fring} and \cite{E2Fring2}, respectively. Both Hamiltonians are
expressed in terms of the $E_{2}$-basis operators $u$, $v$ and $J$ that obey
the commutation relations 
\begin{equation}
\left[ u,J\right] =iv,\qquad \left[ v,J\right] =-iu,\qquad \left[ u,v\right]
=0.
\end{equation}%
Except for $\mathcal{H}_{E_{2}}^{(0)}$ at $N=1/4$, both Hamiltonians are
non-Hermitian, but respect the anti-linear symmetry \cite{EW} $\mathcal{PT}%
_{3}:$ $J\rightarrow J$, $u\rightarrow v$, $v\rightarrow u$, $i\rightarrow
-i $ as defined in \cite{DFM}. For the particular representation $%
J:=-i\partial _{\theta }$, $u:=\sin \theta $ $v:=\cos \theta $ the $\mathcal{%
PT}_{3}$-symmetry is simply $\mathcal{PT}_{3}:\theta \rightarrow \pi
/2-\theta $, $i\rightarrow -i$, such that the invariant vector spaces over $%
\mathbb{R}$ were defined as 
\begin{eqnarray}
V_{n}^{s}(\phi _{0}) &:&=\limfunc{span}\left\{ \left. \phi _{0}\left[ \sin
(2\theta ),i\sin (4\theta ),\ldots ,i^{n+1}\sin (2n\theta )\right]
\right\vert \theta \in \mathbb{R},\mathcal{PT}_{3}(\phi _{0})=\phi _{0}\in
L\right\} ,~~~~~~  \label{q1} \\
V_{n}^{c}(\phi _{0}) &:&=\limfunc{span}\left\{ \left. \phi _{0}\left[
1,i\cos (2\theta ),\ldots ,i^{n}\cos (2n\theta )\right] \right\vert \theta
\in \mathbb{R},\mathcal{PT}_{3}(\phi _{0})=\phi _{0}\in L\right\} .
\label{q2}
\end{eqnarray}%
In order to construct Hamiltonians that preserve the flag structure one
needs to identify the action of the $E_{2}$-basis operators and its
combinations on these spaces as explained in more detail in \cite{E2Fring}.
The behaviour found allowed to identify the Hamiltonians $\mathcal{H}%
_{E_{2}}^{(1)}$ and $\mathcal{H}_{E_{2}}^{(0)}$ in (\ref{H1}) and (\ref{H0})
as quasi-exactly solvable. The general structure suggests that there might
be a master Hamiltonian that unifies the above Hamiltonians into one
preserving the quasi-exact solvability. We demonstrate here that this is
possible and study the properties of that model.

Thus we introduce the new Hamiltonian%
\begin{equation}
\mathcal{H}(N,\zeta ,\lambda )=J^{2}+2(1-\lambda )\zeta uvJ+\lambda \zeta
^{2}(u^{2}-v^{2})^{2}+2i\zeta N(u^{2}-v^{2}),\qquad \lambda ,\zeta ,N\in 
\mathbb{R},  \label{newH}
\end{equation}%
and demonstrate explicitly that it is\ indeed $E_{2}$-quasi-exactly
solvable. First we observe that $\mathcal{H}(N,\zeta ,\lambda )$
interpolates between the two models in (\ref{H1}) and (\ref{H0}) by varying $%
\lambda $, since%
\begin{equation}
\lim_{\lambda \rightarrow 1}\mathcal{H}(N,\zeta ,\lambda )=\mathcal{H}%
_{E_{2}}^{(1)}\quad \quad \text{and\quad \quad }\lim_{\lambda \rightarrow 0}%
\mathcal{H}(2N,\zeta /2,\lambda )=\mathcal{H}_{E_{2}}^{(0)}.
\end{equation}%
Furthermore, $\mathcal{H}(N,\zeta ,\lambda )$ reduces to the complex Mathieu
Hamiltonian in the double scaling limit $\lim_{N\rightarrow \infty ,\zeta
\rightarrow 0}\mathcal{H}(N,\zeta ,\lambda )=\mathcal{H}_{\text{Mat}%
}=J^{2}+2ig(u^{2}-v^{2})$ for $g:=N\zeta <\infty $. We also note that $%
\mathcal{H}^{\dagger }(N,\zeta ,\lambda )=\mathcal{H}(1-\lambda -N,\zeta
,\lambda )$, which implies that $\mathcal{H}(N,\zeta ,\lambda )$ is
non-Hermitian unless $2N=1-\lambda $, with free coupling constant $\zeta \in 
\mathbb{R}$.

Given the structure for the vector spaces in (\ref{q1}) and (\ref{q2}) we
now make the following Ans\"{a}tze for the two fundamental solutions of the
corresponding Schr\"{o}dinger equation $\mathcal{H}_{N}\psi _{N}=E\psi _{N}$ 
\begin{equation}
\psi _{N}^{c}(\theta )=\phi _{0}\sum_{n=0}^{\infty }i^{n}c_{n}P_{n}(E)\cos
(2n\theta ),\quad \text{and\quad }\psi _{N}^{s}(\theta )=\phi
_{0}\sum_{n=0}^{\infty }i^{n+1}c_{n}Q_{n}(E)\sin (2n\theta ),  \label{FS}
\end{equation}%
where the $\mathcal{PT}_{3}$-symmetric ground state is taken to be $\phi
_{0}=e^{\frac{i}{2}\zeta \cos (2\theta )}$ and the constant $c_{n}$ is $%
c_{n}=1/\zeta ^{n}(N+\lambda )(1+\lambda )^{n-1}\left[ (1+N+2\lambda
)/(1+\lambda )\right] _{n-1}$ with $(a)_{n}:=\Gamma \left( a+n\right)
/\Gamma \left( a\right) $ denoting the Pochhammer symbol. The constants are
chosen conveniently in order to ensure the simplicity of the to be
determined $n$-th and $(n-1)$-th order polynomials $P_{n}(E)$, $Q_{n}(E)$ in
the energies $E$, respectively. Upon substitution into the Schr\"{o}dinger
equation we obtain the three-term recurrence relations%
\begin{eqnarray}
P_{2} &=&(E-\lambda \zeta ^{2}-4)P_{1}+\mathbf{2}\zeta ^{2}\left[ N-1\right] %
\left[ N+\lambda \right] P_{0},  \label{r1} \\
P_{n+1} &=&(E-\lambda \zeta ^{2}-4n^{2})P_{n}+\zeta ^{2}\left[ N+n\lambda
+(n-1)\right] \left[ N-(n-1)\lambda -n\right] P_{n-1},  \label{r2} \\
Q_{2} &=&(E-4-\lambda \zeta ^{2})Q_{1},  \label{r3} \\
Q_{m+1} &=&(E-\lambda \zeta ^{2}-4m^{2})Q_{m}+\zeta ^{2}\left[ N+m\lambda
+(m-1)\right] \left[ N-(m-1)\lambda -m\right] Q_{m-1},~~~~~  \label{r4}
\end{eqnarray}%
for $n=0,2,\ldots $ and for $m=2,3,4,\ldots $ Note that a more generic
Ansatz for the unifying model involving two independent coupling constants $%
\mu $, $\lambda $ in the terms $\mu \zeta uvJ+\lambda \zeta
^{2}(u^{2}-v^{2})^{2}$ leads to a four term recurrence relation in which the
highest term is always proportional to $\mu +2\lambda -2$. Thus taking this
term to zero with the appropriate choice for $\mu $ reduces this to the
desired three term relations that may be solved in complete generality as
outlined in \cite{E2Fring}. The lowest order polynomials are easily computed
in a recursive way. Taking $P_{0}=1$ we obtain 
\begin{eqnarray}
P_{1} &=&E-\lambda \zeta ^{2}, \\
P_{2} &=&\lambda ^{2}\zeta ^{4}+2\zeta ^{2}\left[ \lambda -\lambda
E+N(\lambda +N-1)\right] +(E-4)E,  \notag \\
P_{3} &=&-\lambda ^{3}\zeta ^{6}+\lambda \zeta ^{4}\left( \lambda (2\lambda
+3E-13)-3N^{2}-3(\lambda -1)N+2\right) +(E-16)(E-4)E  \notag \\
&&-\zeta ^{2}\left[ 3\lambda E^{2}+E\left( 2\lambda ^{2}-3N^{2}-3\lambda
(N+11)+3N+2\right) +32(\lambda +N(\lambda +N-1))\right] ,  \notag
\end{eqnarray}%
and likewise with $Q_{1}=1$ we compute 
\begin{eqnarray}
Q_{2} &=&E-4-\lambda \zeta ^{2}, \\
Q_{3} &=&\lambda ^{2}\zeta ^{4}+\zeta ^{2}\left[ \lambda (15-2\lambda
-2E)+N^{2}+(\lambda -1)N-2\right] +(E-16)(E-4),  \notag \\
Q_{4} &=&-\lambda ^{3}\zeta ^{6}+\lambda \zeta ^{4}\left[ 8+\lambda
(8\lambda +3\text{$E$}-38)-2N^{2}-2(\lambda -1)N\right] +(\text{$E$}-36)(%
\text{$E$}-16)(\text{$E$}-4)  \notag \\
&&+\zeta ^{2}\left[ -8\left( -12\lambda ^{2}+69\lambda +5\lambda
N+5(N-1)N-12\right) \right]  \notag \\
&&+\zeta ^{2}\left[ -3\lambda \text{$E$}^{2}+2\text{$E$}\left( (47-4\lambda
)\lambda +N^{2}+(\lambda -1)N-4\right) \right] .  \notag
\end{eqnarray}%
In both cases we observe the typical feature for quasi-exactly solvable
systems that the three term relation can be reset to a two-term relation at
a certain level. This is due to the fact that in (\ref{r2}) and (\ref{r4})
the last term vanishes when $m=n=\hat{n}=-(1+N)/(1+\lambda )$ or $m=n=\tilde{%
n}=(\lambda +N)/(1+\lambda )$. Thus when taking $N=\tilde{n}+(\tilde{n}%
-1)\lambda $ we find the typical factorization%
\begin{equation}
P_{\tilde{n}+\ell }=P_{\tilde{n}}R_{\ell }\qquad \text{and\qquad }Q_{\tilde{n%
}+\ell }=Q_{\tilde{n}}R_{\ell }.
\end{equation}%
The first solutions for the factor $R_{\ell }$ are easily found from (\ref%
{r2}) and (\ref{r4}) to 
\begin{eqnarray}
R_{1} &=&E-4\tilde{n}^{2}-\lambda \zeta ^{2}, \\
R_{2} &=&(E-4\tilde{n}^{2}-\lambda \zeta ^{2})(E-4(\tilde{n}+1)^{2}-\lambda
\zeta ^{2})-2\tilde{n}(1+\lambda )^{2}\zeta ^{2}.
\end{eqnarray}

Next we compute the energy eigenvalues $E_{\tilde{n}}$ from the constraints $%
P_{\tilde{n}}(E)=0$ and $Q_{\tilde{n}}(E)=0$ for the lowest values of $N$.
For the solutions related to the even fundamental solution in (\ref{FS}) we
find 
\begin{eqnarray}
N &=&1:\qquad \ \ \ \ \ \ \ \ \ E_{1}^{c}=\lambda \zeta ^{2}, \\
N &=&2+\lambda :\qquad \ \ E_{2}^{c,\pm }=2+\lambda \zeta ^{2}\pm 2\sqrt{%
1-(1+\lambda )^{2}\zeta ^{2}}, \\
N &=&3+2\lambda :\qquad E_{3}^{c,\ell }=\frac{20}{3}+\lambda \zeta ^{2}+%
\frac{4\hat{\Omega}}{3}e^{\frac{i\pi \ell }{3}}+\frac{1}{3}\left[
52-12(1+\lambda )^{2}\zeta ^{2}\right] e^{-\frac{i\pi \ell }{3}}\hat{\Omega}%
^{-1},\qquad ~~  \label{E5}
\end{eqnarray}%
with $\hat{\Omega}^{3}:=35+18(\lambda +1)^{2}\zeta ^{2}+\sqrt{\left[
3(\lambda +1)^{2}\zeta ^{2}-13\right] ^{3}+\left[ 18(\lambda +1)^{2}\zeta
^{2}+35\right] ^{2}}$, $\ell =0,\pm 2$.

For the solutions related to the odd fundamental solution in (\ref{FS}) we
obtain 
\begin{eqnarray}
N &=&2+\lambda :\qquad \ \ \ E_{2}^{s}=4+\lambda \zeta ^{2}, \\
N &=&3+2\lambda :\qquad \ E_{3}^{s,\pm }=10+\zeta ^{2}\lambda \pm 2\sqrt{%
9-(\lambda +1)^{2}\zeta ^{2}}, \\
N &=&4+3\lambda :\qquad E_{4}^{s,\ell }=\frac{56}{3}+\lambda \zeta ^{2}+%
\frac{4\Omega }{3}e^{\frac{i\pi \ell }{3}}+\frac{1}{3}\left[
196-12(1+\lambda )^{2}\zeta ^{2}\right] e^{-\frac{i\pi \ell }{3}}\Omega
^{-1},\qquad ~~  \label{E7}
\end{eqnarray}%
with $\Omega ^{3}:=143+18\zeta ^{2}(\lambda +1)^{2}+\sqrt{\left( 3\zeta
^{2}(\lambda +1)^{2}-49\right) ^{3}+\left( 18\zeta ^{2}(\lambda
+1)^{2}+143\right) ^{2}}$, $\ell =0,\pm 2$. Solutions for higher order may
of course also be obtained, but are rather lengthy and therefore not
reported here.

\section{Exceptional points and their vicinities}

The special point in parameter space where two real energy eigenvalues
viewed as functions of the coupling constants merge and subsequently split
into a complex conjugate pair is usually referred to as exceptional point 
\cite{Kato,HeissEx,IngridEx,IngridUwe}. In our system these points can be
computed in an explicit simple and straightforward manner. Using that by
definition the discriminant $\Delta $ equals the product of the squares of
the differences of all energy eigenvalues $E_{i}$ for $1\leq i\leq n$, i.e. $%
\Delta =\prod\nolimits_{1\leq i<j\leq n}(E_{i}-E_{j})^{2}$ one obtains the
exceptional points from the real zeros of $\Delta (E)$. For practical
purposes one may also exploit the fact \cite{E2Fring}, that the discriminant
equals the determinant of the Sylvester matrix. This viewpoint has the
advantage that it does not require the computation of all the eigenvalues
and is more efficient when the sole purpose is to find the exceptional
points. Thus in our case we have to find the real zeros of the discriminants 
$\Delta _{\tilde{n}}^{c}$ and $\Delta _{\tilde{n}}^{s}$ for the polynomials $%
P_{\tilde{n}}(E)$ and $Q_{\tilde{n}}(E)$, respectively. Extracting overall
constant factors $\kappa $ as $\Delta =\kappa \tilde{\Delta}$, that do not
contribute to the zeros, we obtain for the lowest values of $\tilde{n}$ 
\begin{eqnarray}
\tilde{\Delta}_{2}^{c} &=&\hat{\zeta}^{2}-1,  \label{d1} \\
\tilde{\Delta}_{3}^{s} &=&\hat{\zeta}^{2}-9,  \notag \\
\tilde{\Delta}_{3}^{c} &=&\hat{\zeta}^{6}-\hat{\zeta}^{4}+103\hat{\zeta}%
^{2}-36,  \notag \\
\tilde{\Delta}_{4}^{s} &=&\hat{\zeta}^{6}-37\hat{\zeta}^{4}+991\hat{\zeta}%
^{2}-3600,  \notag \\
\tilde{\Delta}_{4}^{c} &=&\hat{\zeta}^{12}+2\hat{\zeta}^{10}+385\hat{\zeta}%
^{8}-33120\hat{\zeta}^{6}+16128\hat{\zeta}^{4}-732276\hat{\zeta}^{2}+129600,
\notag \\
\tilde{\Delta}_{5}^{s} &=&\hat{\zeta}^{12}-94\hat{\zeta}^{10}+7041\hat{\zeta}%
^{8}-381600\hat{\zeta}^{6}+6645600\hat{\zeta}^{4}-78318900\hat{\zeta}%
^{2}+158760000,  \notag  \label{d6}
\end{eqnarray}%
where we abbreviated $\hat{\zeta}:=\zeta (1+\lambda )$.

There exist many detailed studies about the structures in the coupling
constant space in the vicinity of the exceptional points \cite%
{HeissHar,Mehri,RotterRev,HeissRev,HeissWunner}. It is evident that when
tracing a complex energy eigenvalue $E$ as functions of the coupling
constants, $\lambda $ or $\zeta $ in our case, the corresponding path in the
energy plane will inevitably pass through various Riemann sheets due to the
branch cut structure. As a consequence one naturally generates eigenvalue
loops that stretch over several Riemann sheets. This phenomenon is well
studied for a large number of models and we demonstrate here that it also
occurs in quasi-exactly solvable models. The basic principle can be
demonstrated with the square root singularity occurring in $E_{2}^{c,\pm }$
with branch cuts from $(-\infty ,-1-1/\zeta )$ and $(1/\zeta -1,\infty )$.
The energy loops are generated by computing $E_{2}^{c,\pm }(\lambda =\tilde{%
\lambda}+\rho e^{i\pi \phi },\zeta )$ for some fixed values of $\zeta $,
center $\tilde{\lambda}$ and the radius $\rho $ in the $\lambda $-plane as
functions of $\phi $ as illustrated in figure \ref{loop}(a) and (b). In
panel (a) we simply trace the energy around a point in parameter space that
leads to two real eigenvalues. For a small radius ones reaches the starting
point by encircling $\tilde{\lambda}$ just once. However, when the radius is
increased one needs to surround $\tilde{\lambda}$ twice to reach the
starting point and when the radius is increased even further one only needs
to surround $\tilde{\lambda}$ once switching, however, between both energy
eigenvalues.

\FIGURE{\epsfig{file=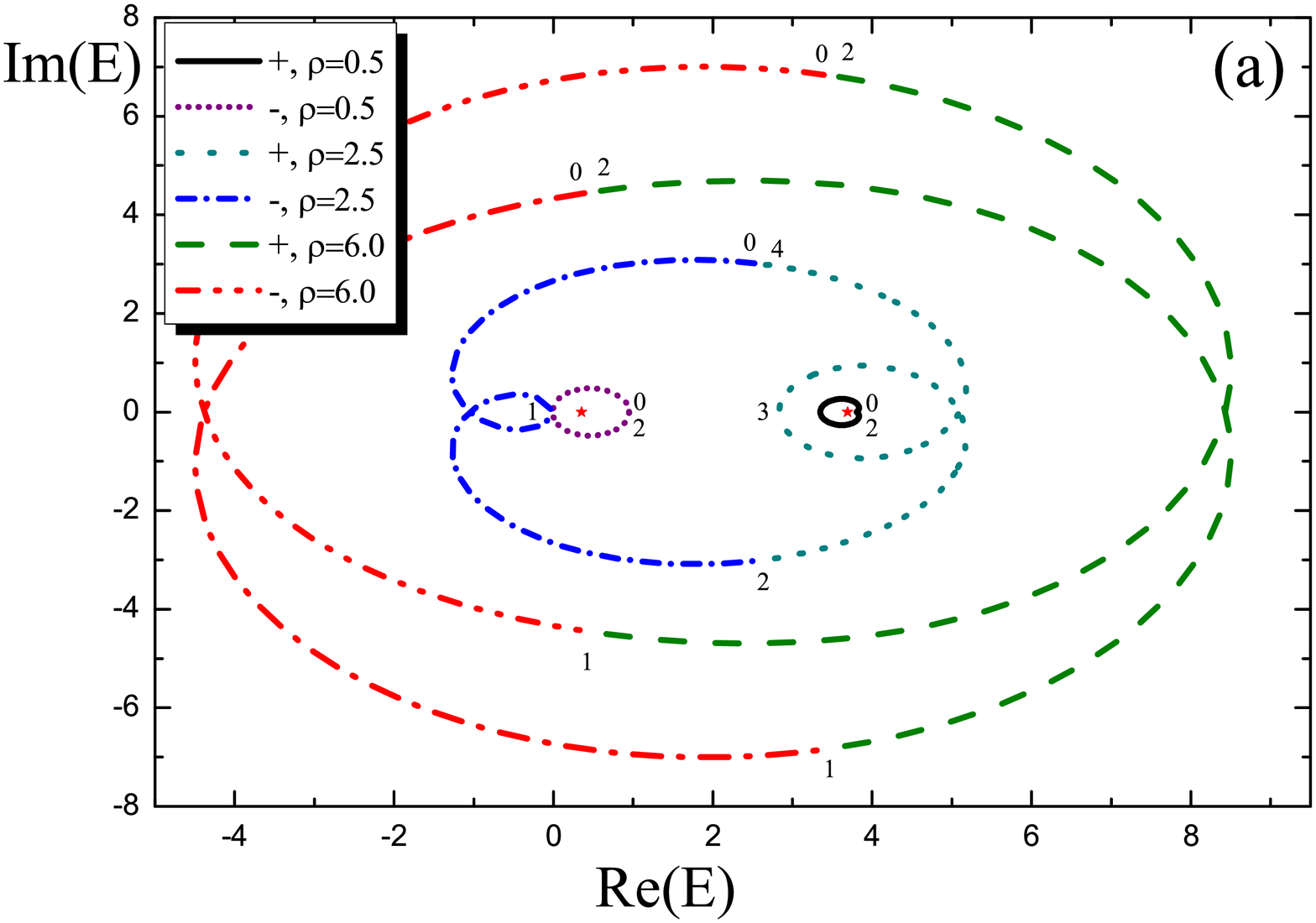,width=7.2cm} \epsfig{file=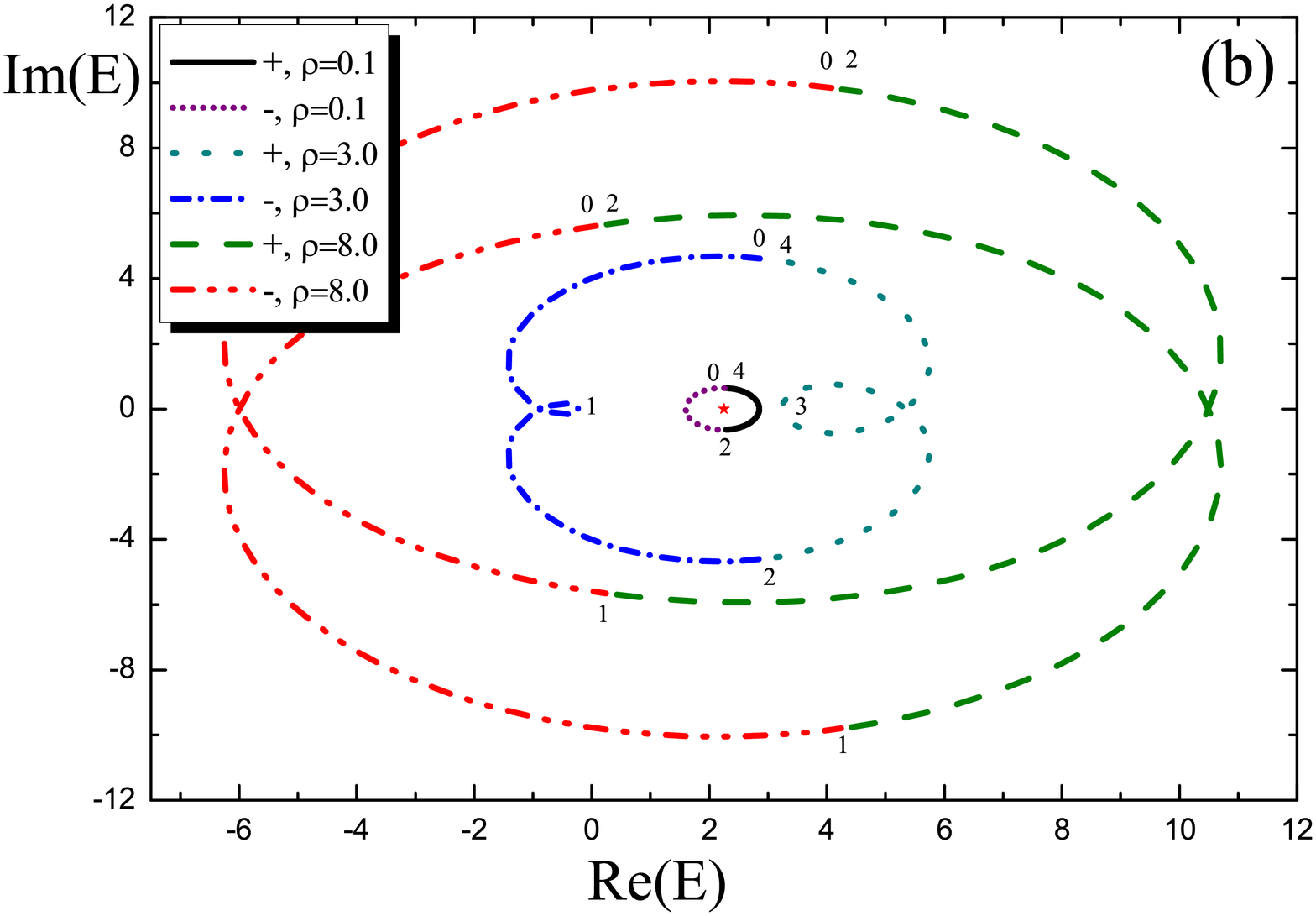,width=7.2cm} 
        \caption{Energy eigenvalue loops $E_{2}^{c,\pm }(\tilde{\lambda}+\rho e^{i\pi \phi },\zeta )
$ around two real eigenvalues panel (a) and around an exceptional point panel (b) as functions of $\phi $, indicated by the numbers on the loops, for fixed
value of $\zeta =1/2$ at $\tilde{\lambda}=1/10$ in (a) and $\tilde{\lambda}=1
$ in (b). The energy eigenvalues for $\rho =0$ are distinct in panel (a) as $E_{2}^{c,-}=0.35$, $E_{2}^{c,+}=3.70$ and coalesce to an exceptional point
in panel (b) as $E_{2}^{c,-}=E_{2}^{c,+}=9/4$. }
        \label{loop}} 
\FIGURE{\epsfig{file=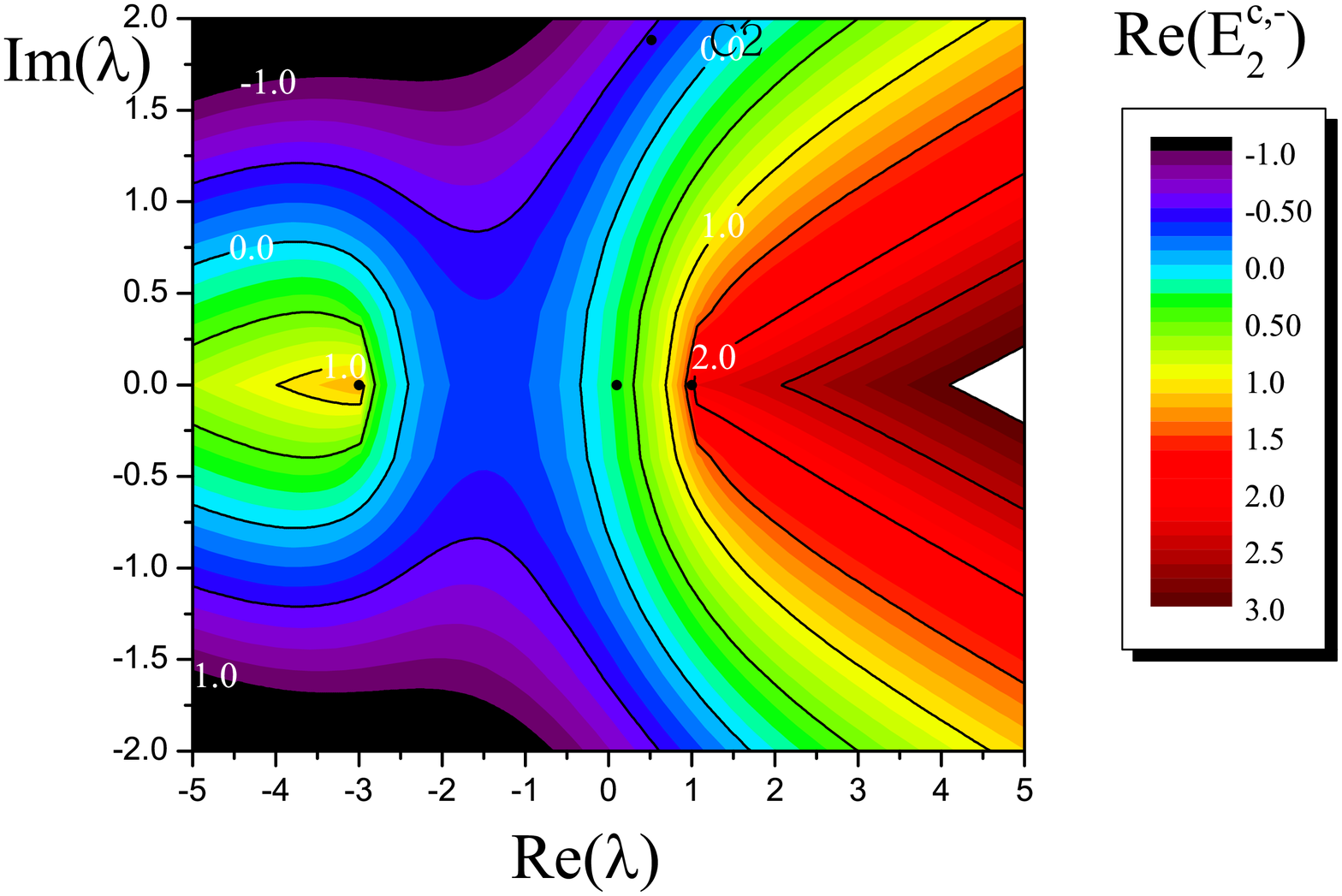,width=7.2cm} \epsfig{file=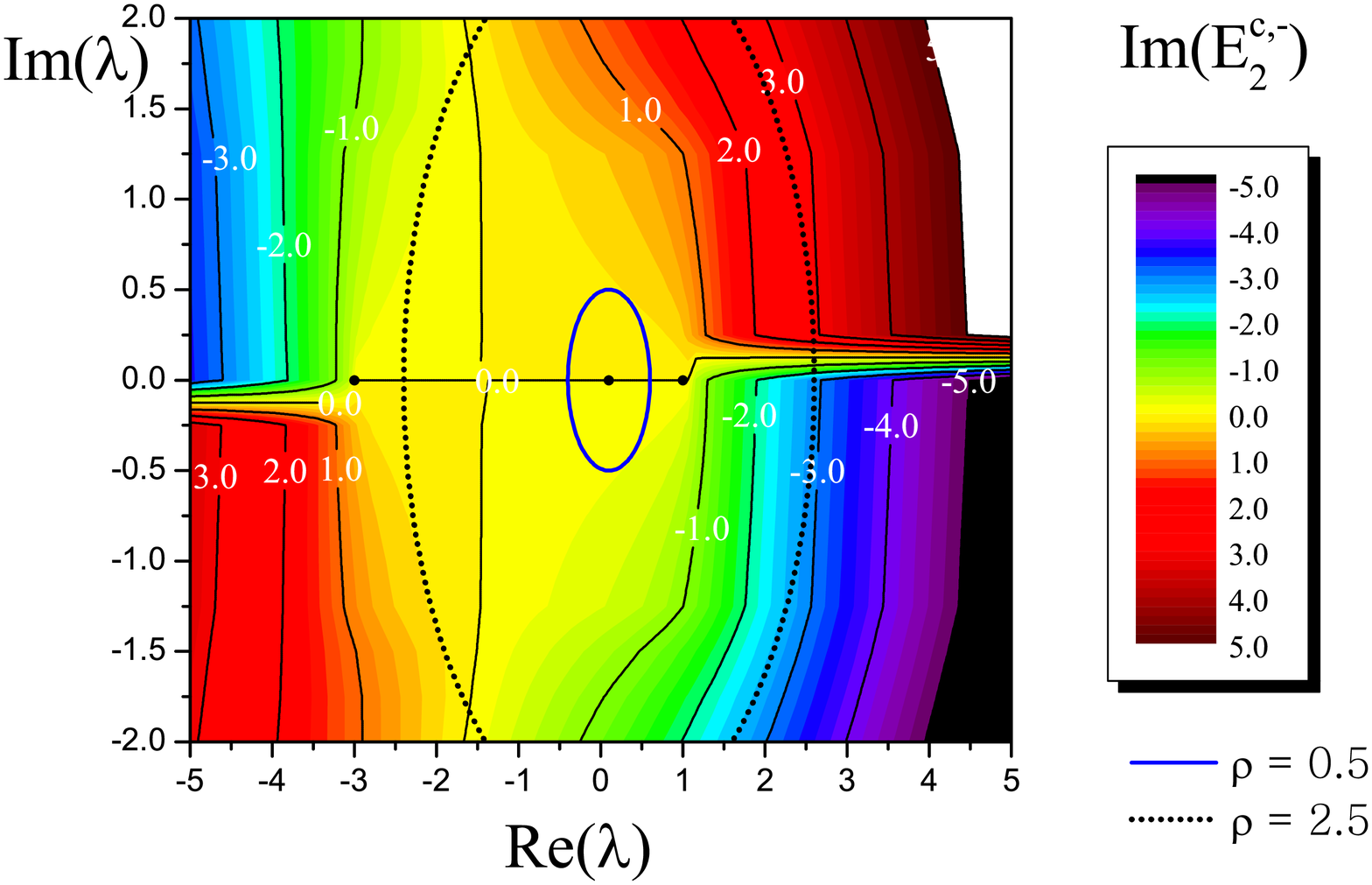,width=7.2cm} 
        \epsfig{file=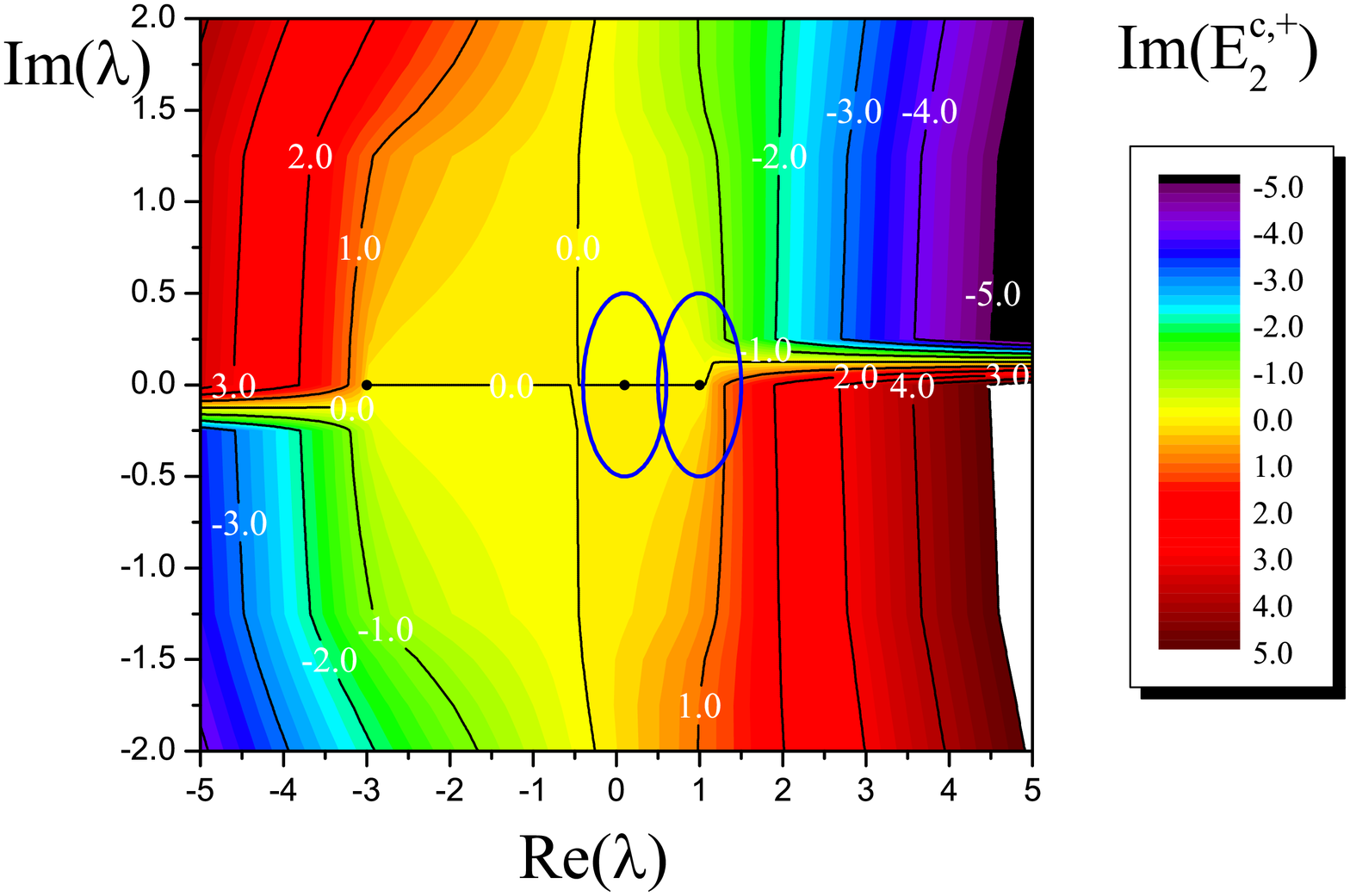,width=7.2cm} \epsfig{file=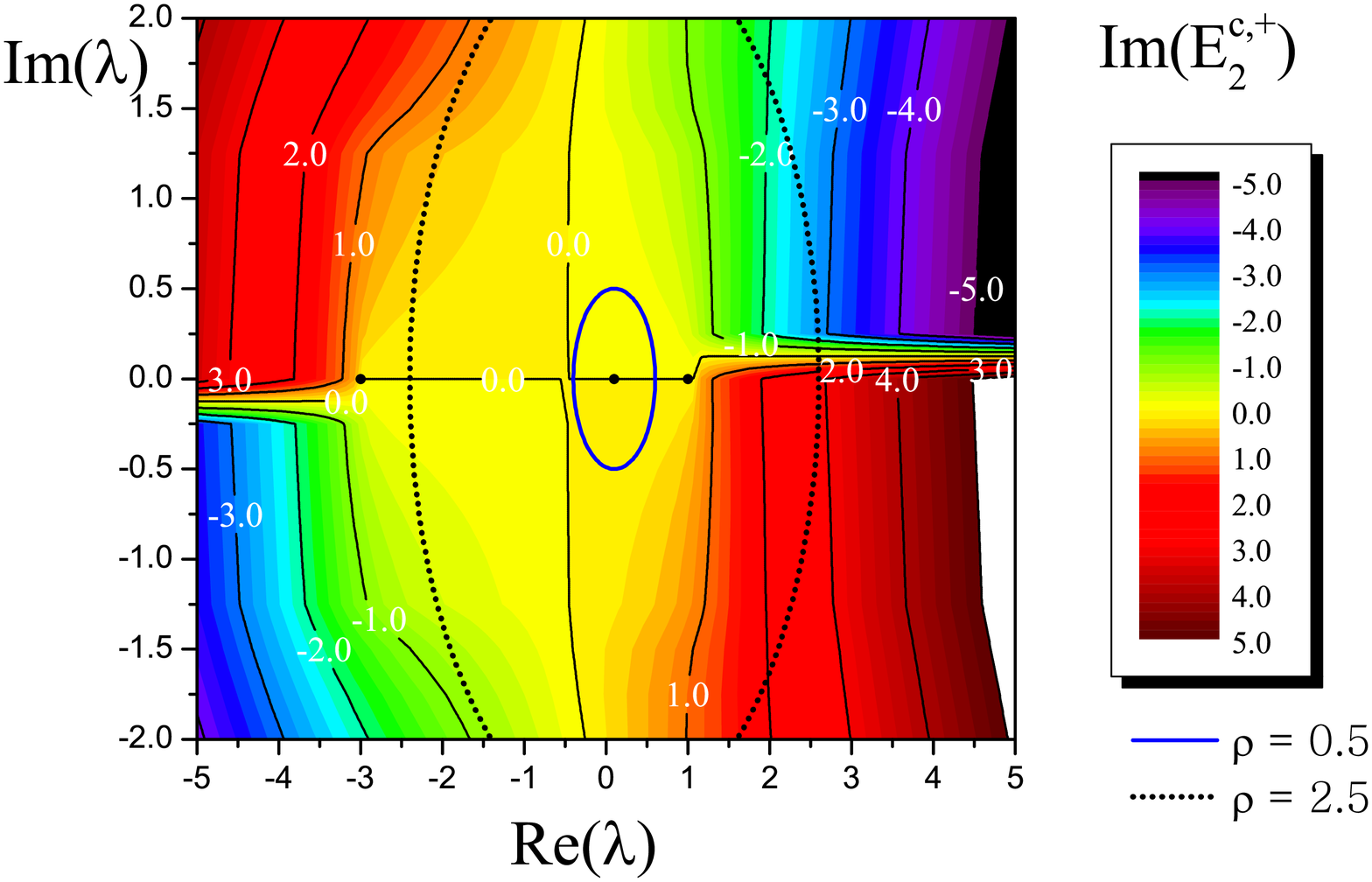,width=7.2cm} 
        \caption{Energy levels and branch cut structure for $E_{2}^{c,\pm }$ for fixed $\zeta=1/2$ as functions of $\lambda$. The branch cuts extend to the left and right from the exceptional points 
$\left( -\infty ,-3\right) $ and $(1,\infty )$.}
        \label{EM}}

Essentially this structure survives when the two eigenvalues merge into an
exceptional point. However, since the exceptional point is a branch point we
no longer have the option for a closed loop around it produced from only one
energy eigenvalue as seen in figure \ref{loop}(b).

This behaviour is easily understood from the structure of the branch cuts as
depicted in figure \ref{EM}. Whereas for small radii it is possible to
encircle for instance the point $\tilde{\lambda}=1/10$ without crossing any
branch cut, this is not possible when encircling the exceptional point at $%
\tilde{\lambda}=1$ where we have to analytically continue from $E_{2}^{c,-}$
to $E_{2}^{c,+}$ when crossing a cut. This structure is the same for
intermediate radii. For large radii we cross the first cut already at a half
circle turn, such that one returns back to the original value already after
one complete turn.

When more eigenvalues are present the structure will be more intricate.
Considering for instance a scenario with four eigenvalues in the form of two
complex conjugate eigenvalues and an exceptional point, see figure \ref%
{loopvier}(a), we need to perform again at least two turns in the $\lambda $%
-plane in order to return to the initial position for the energy loops when
surrounding an exceptional point. The two complex conjugate eigenvalues may
be enclosed with just one turn, albeit we require again different energy
eigenvalues for this. When enlarging the radius the loops will eventually
merge as depicted in figure \ref{loopvier}(b) for a situation with a
degenerate complex eigenvalue and two complex eigenvalues. We observe that
for the given values we have to surround the chosen point at least three
times to obtain a closed energy loop surrounding the indicated centers.

\FIGURE{\epsfig{file=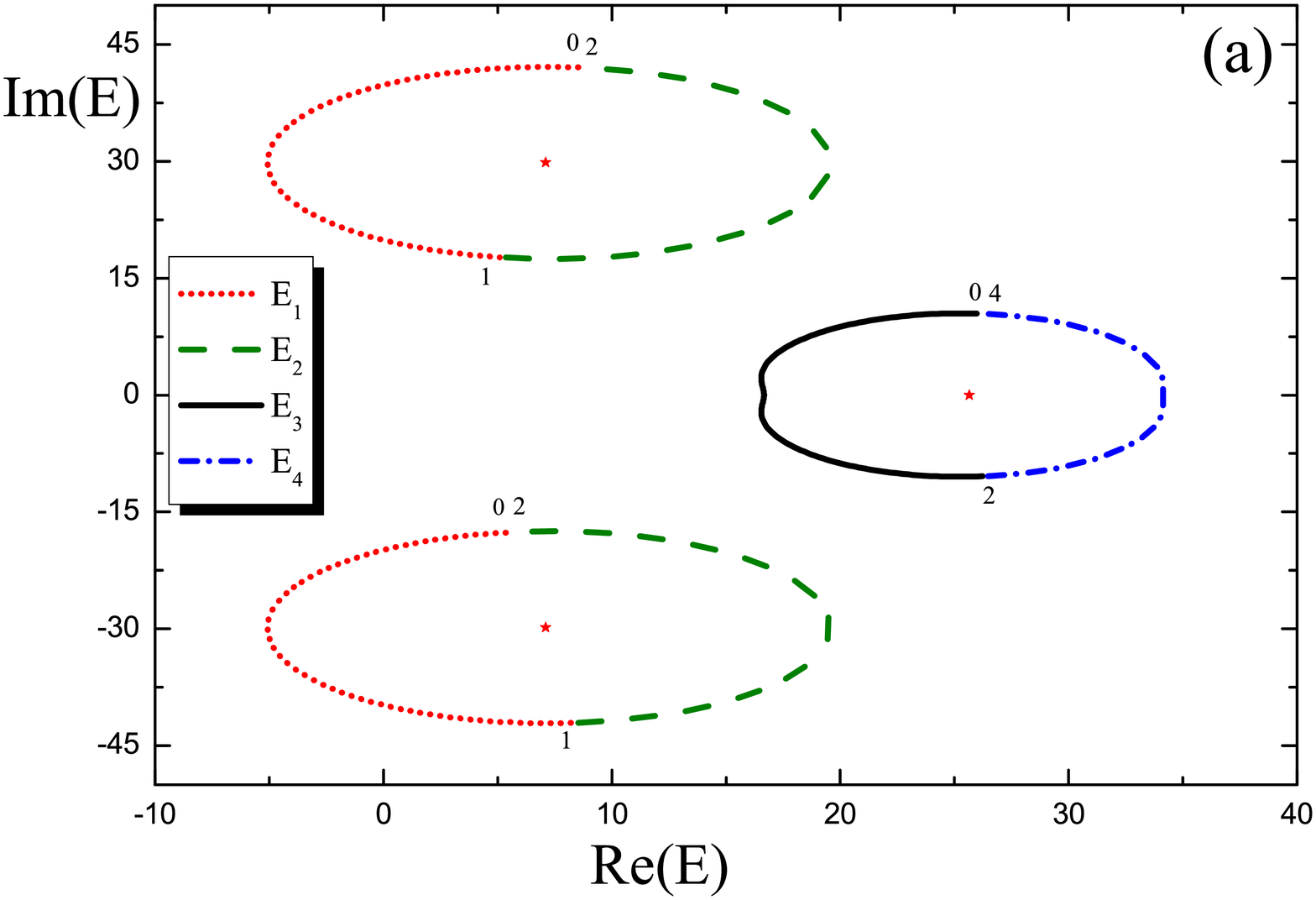,width=7.1cm} \epsfig{file=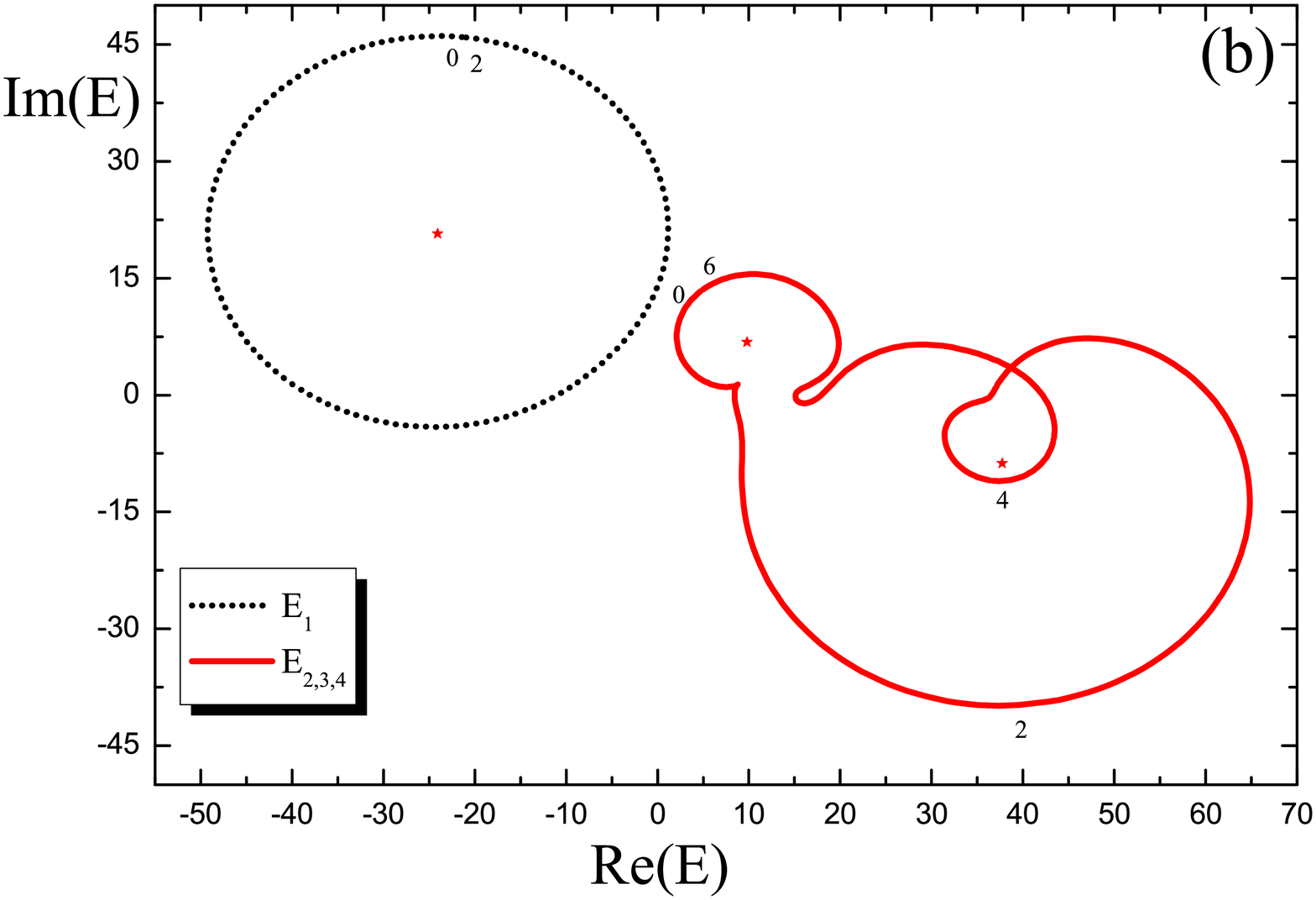,width=7.1cm} 
        \caption{Energy eigenvalues $E_{4}^{c}(\tilde{\lambda}+\rho e^{i\pi \phi },\zeta )
$ as functions of $\phi $, indicated by the numbers on the loops, for fixed
value $\zeta =1/2$ at $\tilde{\lambda}=9.5284$ in (a) and $\tilde{\lambda}=5.2562 +i 9.9526
$ in (b). The energy eigenvalues for $\rho =0$ in panel (a) are $E_{4}^{c,1}=E_{4}^{c,2}=25.6613$, $E_{4}^{c,3}=(E_{4}^{c,4})^*=7.1029 +i 29.8106$ and $E_{4}^{c,1}=E_{4}^{c,2}=37.7449-i 8.7611 $, $E_{4}^{c,3}=9.8103 +i 6.7668$, $E_{4}^{c,4}=-24.0439 +i 20.7081$
in panel (b). The radii are $\rho=4.0$ and $\rho=8.5$ in panel (a) and (b), respectively.}
        \label{loopvier}}
In the same manner as for the simpler scenario one may understand the nature
of these loops from an analysis of the branch cut structure of the energy as
seen in figure \ref{ContE4}. Tracing the indicated radii at $\rho =4.0$ and $%
\rho =8.5$ in figure \ref{ContE4} produces the energy loops in figure \ref%
{loopvier} when properly taking care of the analytic continuation at the
branch cuts.

As discussed earlier the Hamiltonian $\mathcal{H}(N,\zeta ,\lambda )$ has
the interesting property that in the double scaling limit it reduces to the
complex Mathieu equation for which only incomplete information is available,
especially concerning the locations of the exceptional points. In comparison
with the previously analyzed models $\mathcal{H}_{E_{2}}^{(1)}$ in \cite%
{E2Fring} and $\mathcal{H}_{E_{2}}^{(0)}$ in \cite{E2Fring2} we have now the
additional parameter $\lambda $ at our disposal and we may investigate how
the complex Mathieu system is approached. In particular we may address the
question of whether there exists a value $\lambda $ for which this is
optimal. Our numerical results are depicted in figure \ref{opti}. We find a
similar qualitative behaviour for the other exceptional points, which we do
not report here.

Comparing the rate of the approach for different values of $\lambda $ we
conclude that $\mathcal{H}(N,\zeta ,\lambda =1)$ is the best approximation
to the complex Mathieu system for some finite values of $N$.

\FIGURE{\epsfig{file=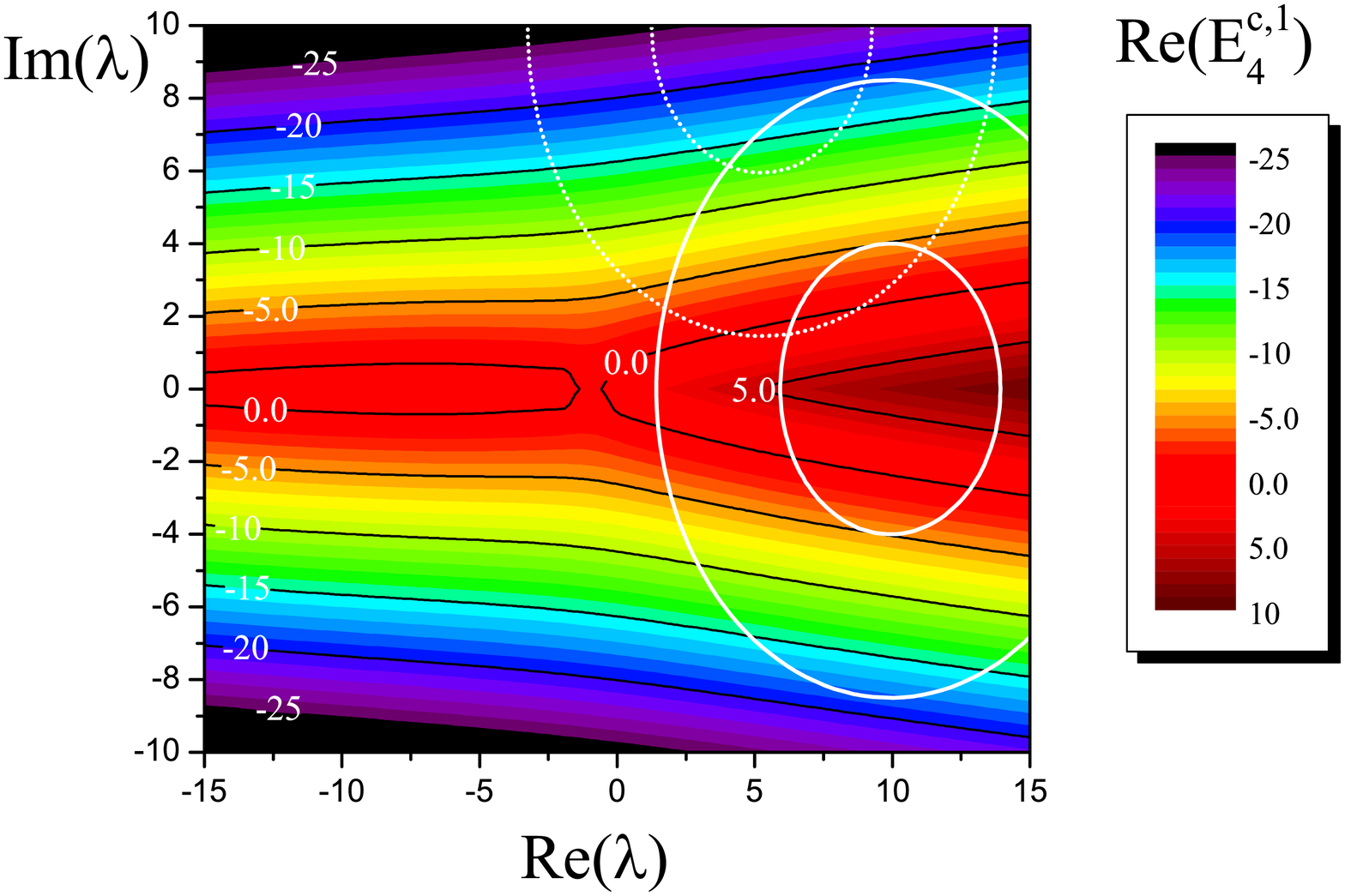,width=7.1cm} \epsfig{file=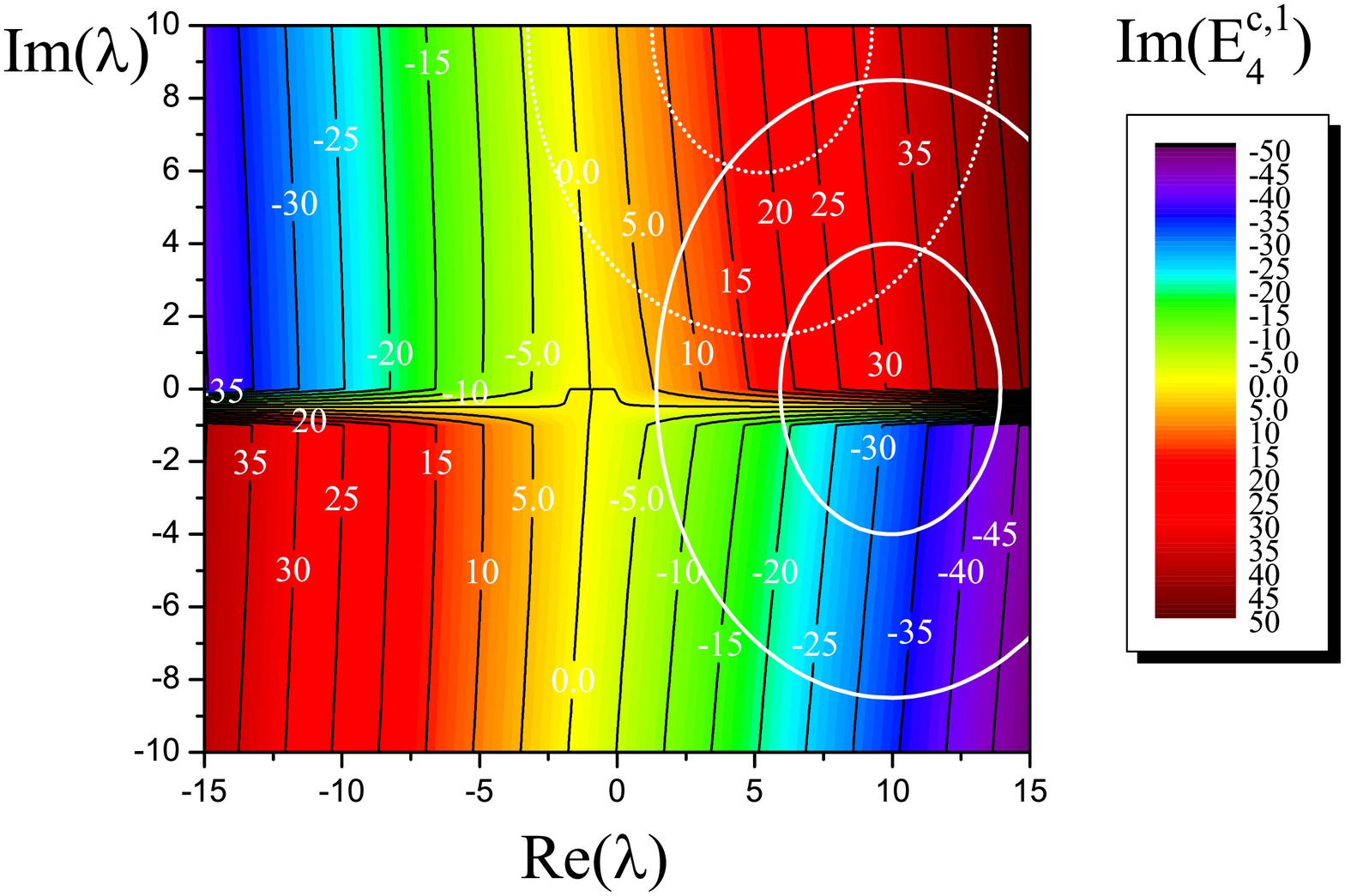,width=7.1cm} 
        \epsfig{file=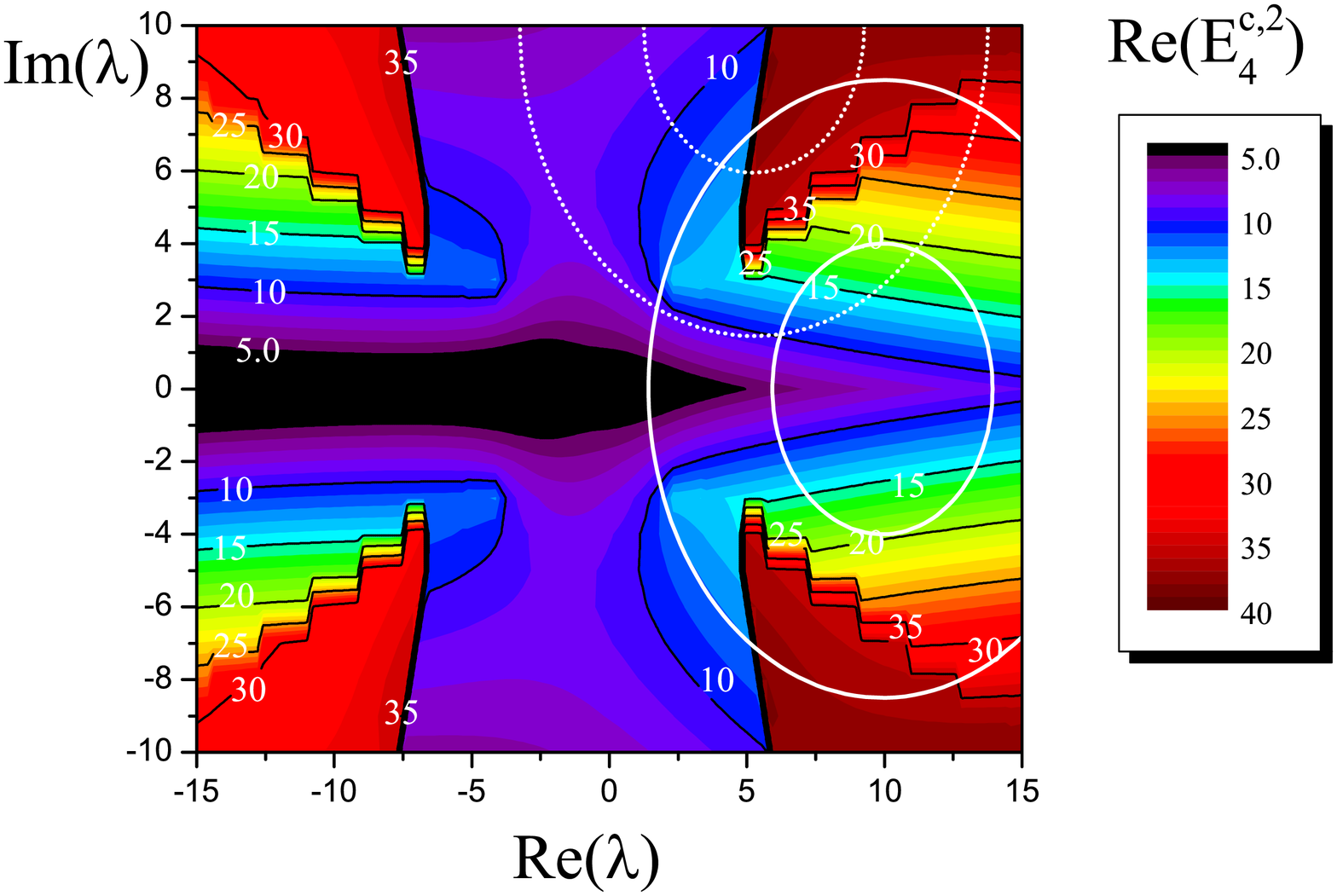,width=7.1cm} \epsfig{file=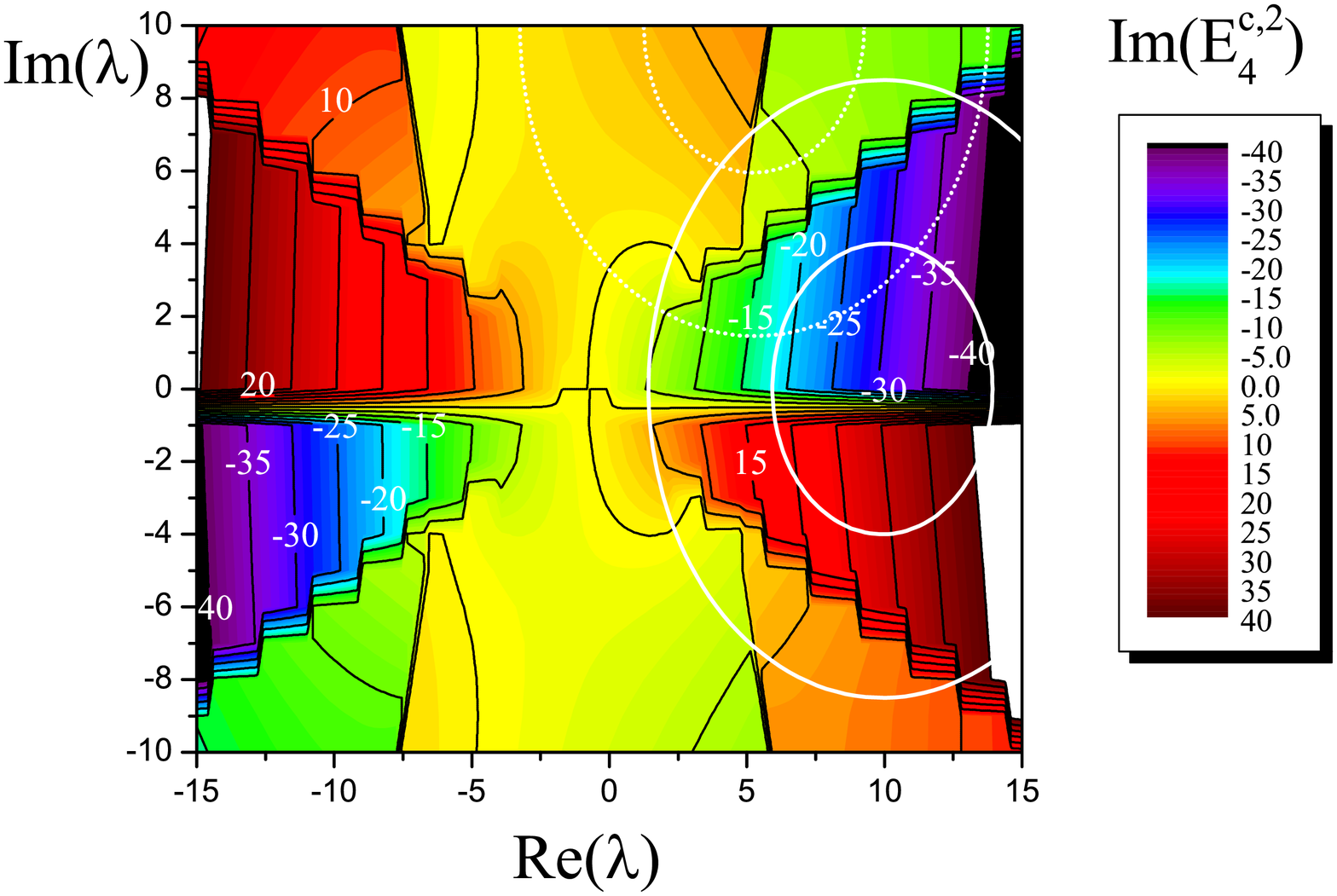,width=7.1cm}
        \epsfig{file=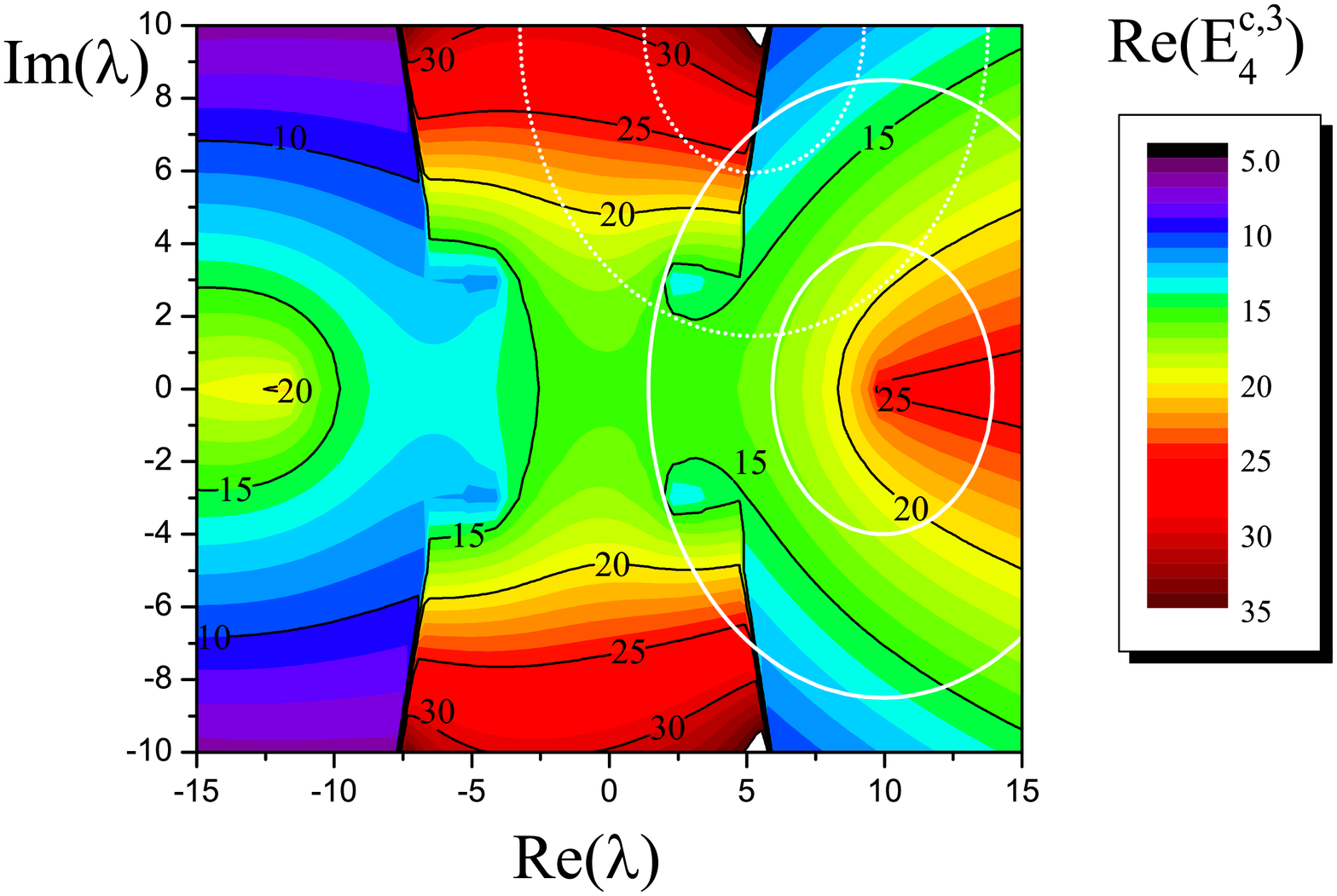,width=7.1cm} \epsfig{file=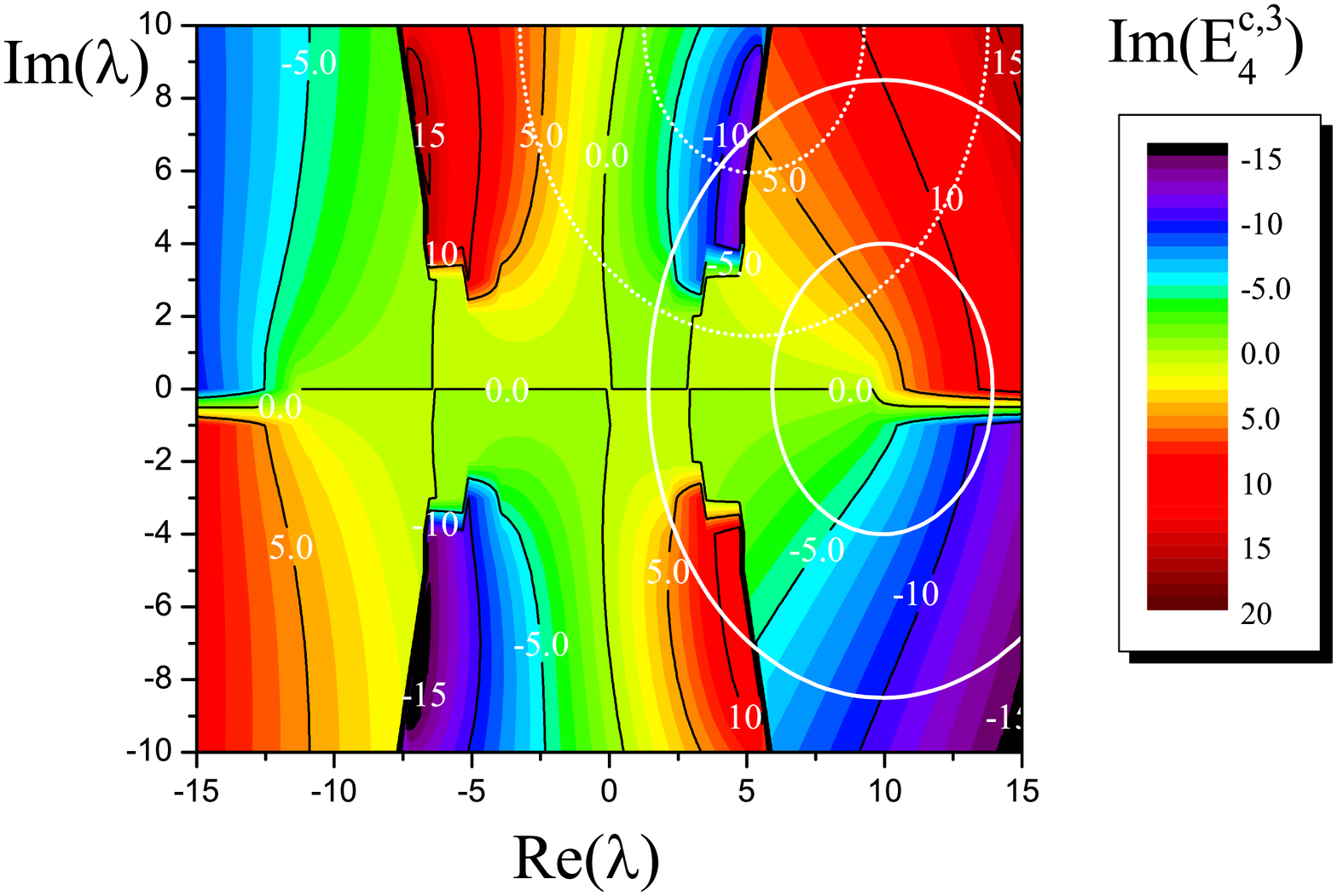,width=7.1cm} 
         \epsfig{file=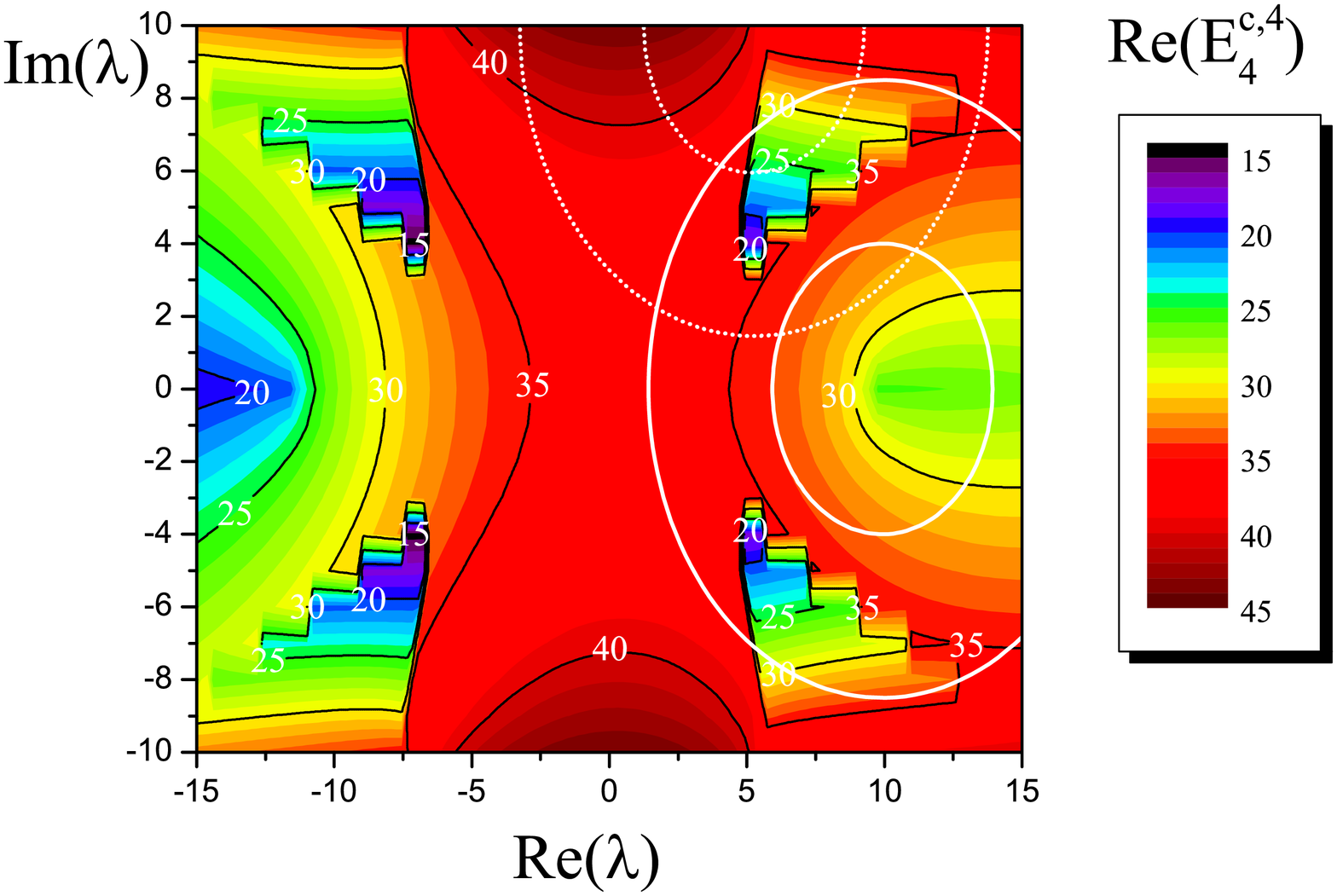,width=7.1cm} \epsfig{file=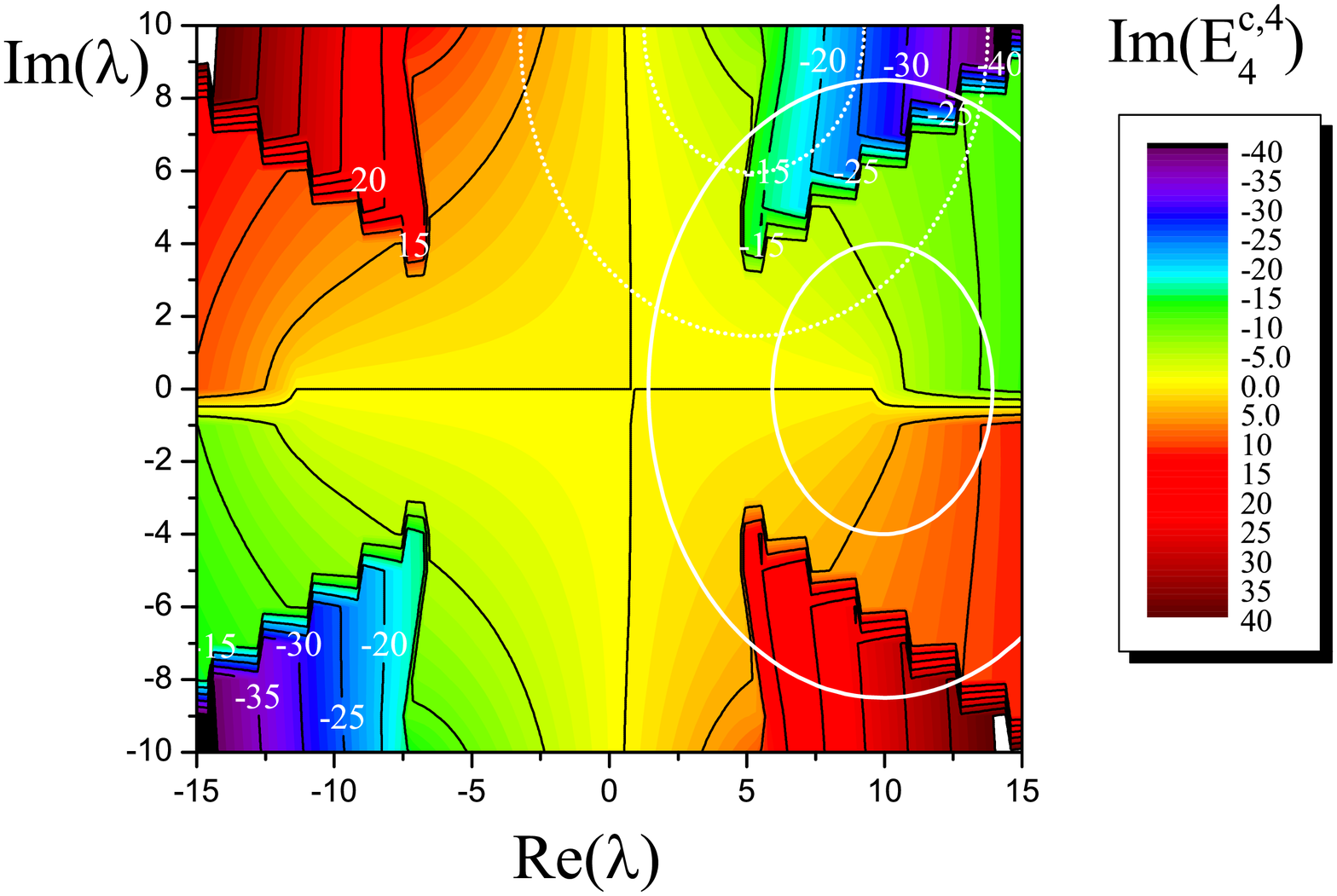,width=7.1cm} 
        \caption{Energy levels and branch cut structure for $E_{2}^{c,1,2,3,4}$ for fixed $\zeta=1/2$ as functions of $\lambda$.}
        \label{ContE4}}

\newpage

If one is exclusively interested in the computation of the exceptional point
it is most efficient to carry out the double scaling limit already for the
three-term relation (\ref{r2}) and (\ref{r4}) as explained in \cite%
{E2Fring,E2Fring2}.

\FIGURE{ \epsfig{file=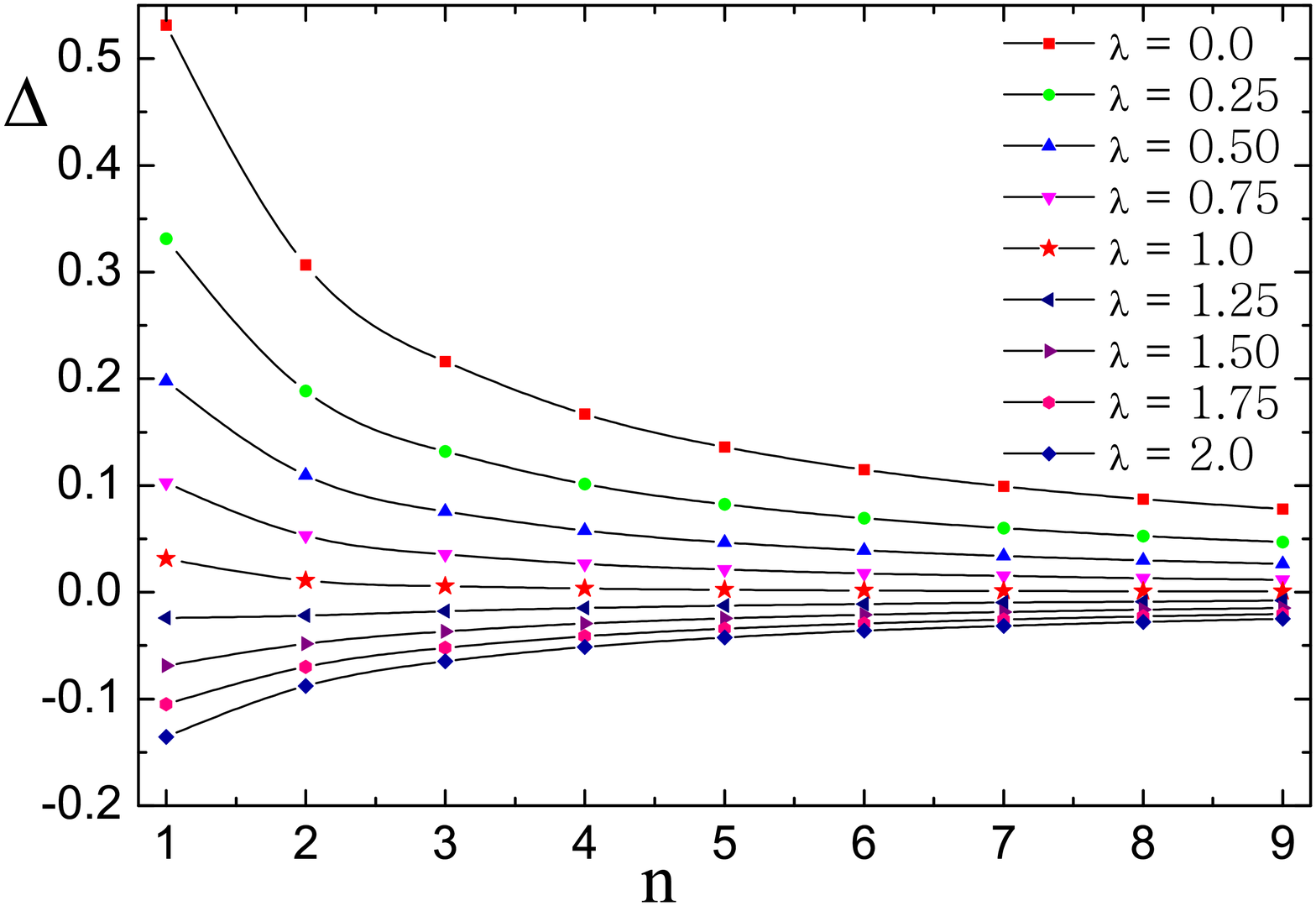,width=10cm} 
        \caption{Double scaling limit of $\lim_{N\rightarrow \infty ,\zeta
\rightarrow 0}\mathcal{H}(N,\zeta ,\lambda )=\mathcal{H}_{\text{Mat}}$ to the smallest exceptional point at $\zeta_M = 1.46877$ with $\Delta(n)= \zeta_0 N(n) -\zeta_M$, $N(n)=(n+1)+n \lambda$ for $n = 1,2,3, \ldots$.}
        \label{opti}}

\section{Weakly orthogonal polynomials}

It is well known from Favard's theorem \cite{Favard,Finkel} that polynomials 
$\Phi _{n}(E)$ constructed from three-term relations in the way mentioned
above possess a norm $N_{n}^{\Phi }$ 
\begin{equation}
\mathcal{L}(\Phi _{n}\Phi _{m})=N_{n}^{\Phi }\delta _{nm}.  \label{norm}
\end{equation}%
defined by the action of a linear functional $\mathcal{L}$ acting on
arbitrary polynomials $p$ in $E$ as%
\begin{equation}
\mathcal{L}(p)=\int\nolimits_{-\infty }^{\infty }p(E)\omega (E)dE,\qquad 
\mathcal{L}(1)=1.  \label{L}
\end{equation}%
This norm may be computed in two alternative ways. The simplest way is to
multiply the three-term relation by $\Phi _{n-1}$ and act subsequently on
the resulting equation with $\mathcal{L}$. Using the property $N_{n}^{\Phi }=%
\mathcal{L}(\Phi _{n}^{2})=\mathcal{L}(E\Phi _{n-1}\Phi _{n})$ together with
(\ref{norm}) then simply yields $N_{n}^{\Phi }=$ $\prod%
\nolimits_{k=1}^{n}b_{k}$, where the\ $b_{k}$ are the negative coefficients
in front of $\Phi _{n-1}$. Whereas the first method simply assumes that the
functional exist the second method goes further and actually provides an
explicit expressions for the measure. As argued in \cite{Krajewska} the
concrete formulae for $\omega (E)$ may be computed from 
\begin{equation}
\omega (E)=\sum\limits_{k=1}^{\ell }\omega _{k}\delta (E-E_{k}),
\end{equation}%
where the energies $E_{k}$ are the $\ell $ roots of the polynomial $\Phi (E)$%
. The $\ell $ constants $\omega _{k}$ can be determined by the $\ell $
equations%
\begin{equation}
\sum\limits_{k=1}^{\ell }\omega _{k}\Phi _{n}(E_{k})=\delta _{n0}\text{,}%
\qquad \text{for }n\in \mathbb{N}_{0}.  \label{om}
\end{equation}%
In our case the integer $\ell $ are determined from $N=\ell +(\ell
-1)\lambda $ and $N=(\ell +1)+\ell \lambda $ for the $P_{\ell }(E)$ and $%
Q_{\ell +1}(E)$, respectively.

Using the first method we obtain

\begin{eqnarray}
N_{n}^{P} &=&2\zeta ^{2n}(1+\lambda )^{2n}\left( \frac{1-N}{1+\lambda }%
\right) _{n}\left( \frac{\lambda +N}{1+\lambda }\right) _{n},~~\quad
n=1,2,3,...  \label{NP} \\
N_{n}^{Q} &=&\frac{1}{2(N+\lambda )(1-N)}N_{n}^{P},\qquad \qquad ~\quad \ \
\ \ \ \ \ \ \ \ \ \ \ n=2,3,4,...  \label{NQ}
\end{eqnarray}%
with $N_{0}^{P}=N_{1}^{Q}=1$. Due to the non-Hermitian nature of the
Hamiltonian this norm is in general not positive definite. For instance for $%
N=4+3\lambda $ we have%
\begin{equation}
N_{0}^{P}=1,~~~N_{1}^{P}=-24\zeta ^{2}(1+\lambda )^{2},~~~N_{2}^{P}=240\zeta
^{4}(1+\lambda )^{4},~~~N_{3}^{P}=-1440\zeta ^{6}(1+\lambda )^{6}.
\end{equation}%
The exception is the class of models where the Hamiltonian becomes
Hermitian, i.e. when $\lambda =1-2N$ holds. For this value of $\lambda $ the
expressions in (\ref{NP}) and (\ref{NQ}) become positive definite%
\begin{equation}
N_{n}^{P}=2^{1+2n}\zeta ^{2n}(N-1)^{2n}\left( \frac{1}{2}\right)
_{n}^{2}=2\zeta ^{2}(N-1)^{2}N_{n}^{Q}.
\end{equation}

Let us now consider the second method and compute explicitly the measure for
a few examples. For $N=2+\lambda $ and $N=3+2\lambda $ we solve (\ref{om})
for the even and odd solutions, respectively, to%
\begin{equation}
\omega _{\pm }^{c}=\frac{1}{2}\pm \frac{1}{2\sqrt{1-(1+\lambda )^{2}\zeta
^{2}}},\quad \text{and\quad }\omega _{\pm }^{s}=\frac{1}{2}\pm \frac{3}{2%
\sqrt{9-(1+\lambda )^{2}\zeta ^{2}}}.
\end{equation}%
Computing now (\ref{norm}) with (\ref{L}) agrees with (\ref{NP}) and (\ref%
{NQ})%
\begin{eqnarray}
N_{0}^{P} &=&\mathcal{L}(P_{0}^{2})=\omega _{+}^{c}+\omega _{-}^{c}=1 \\
N_{1}^{P} &=&\mathcal{L}(P_{1}^{2})=\omega _{+}^{c}\left(
E_{2}^{c,+}-\lambda \zeta ^{2}\right) ^{2}+\omega _{-}^{c}\left(
E_{2}^{c,-}-\lambda \zeta ^{2}\right) ^{2}=-4\hat{\zeta}^{2}, \\
N_{2}^{Q} &=&\mathcal{L}(Q_{2}^{2})=\omega _{+}^{s}\left(
E_{3}^{s,+}-4-\lambda \zeta ^{2}\right) ^{2}+\omega _{-}^{s}\left(
E_{3}^{s,-}-4-\lambda \zeta ^{2}\right) ^{2}=-4\hat{\zeta}^{2}.~~~
\end{eqnarray}%
Similarly we compute for $N=3+2\lambda $ 
\begin{eqnarray}
\omega _{1}^{c} &=&\frac{1}{3}-\frac{\left( 260-60\hat{\zeta}^{2}\right)
\Omega +\left( 3\hat{\zeta}^{2}+4\right) \Omega ^{2}+20\Omega ^{3}}{12\left[
\left( 13-3\hat{\zeta}^{2}\right) ^{2}+\left( 13-3\hat{\zeta}^{2}\right)
\Omega ^{2}+\Omega ^{4}\right] },\qquad \omega _{2}^{c}=\chi _{-2},\qquad
\omega _{3}^{c}=\chi _{2}, \\
\chi _{\ell } &=&\frac{1}{3}+\frac{\left( 3\hat{\zeta}^{2}-20\Omega
+4\right) \left( 1+2e^{\frac{i\pi \ell }{3}}\right) }{36(3\hat{\zeta}%
^{2}+\Omega ^{2}-13)}+\frac{4+3\hat{\zeta}^{2}-20e^{\frac{i\pi \ell }{3}%
}\Omega }{12\left( 1+2e^{\frac{i\pi \ell }{3}}\right) \left( 3\hat{\zeta}%
^{2}-13\right) +\left( 1-e^{\frac{i\pi \ell }{3}}\right) \Omega ^{2}}  \notag
\end{eqnarray}%
and confirm that%
\begin{eqnarray}
N_{0}^{P} &=&\mathcal{L}(P_{0}^{2})=\omega _{1}^{c}+\omega _{2}^{c}+\omega
_{3}^{c}=1,  \label{NN} \\
N_{1}^{P} &=&\mathcal{L}(P_{1}^{2})=\omega
_{1}^{c}P_{1}^{2}(E_{3}^{c,0})+\omega _{2}^{c}P_{1}^{2}(E_{3}^{c,-2})+\omega
_{3}^{c}P_{1}^{2}(E_{3}^{c,2})=-12\hat{\zeta}^{2},  \notag \\
N_{2}^{P} &=&\mathcal{L}(P_{2}^{2})=\omega
_{1}^{c}P_{2}^{2}(E_{3}^{c,0})+\omega _{2}^{c}P_{2}^{2}(E_{3}^{c,-2})+\omega
_{3}^{c}P_{2}^{2}(E_{3}^{c,2})=48\hat{\zeta}^{4}  \notag \\
\mathcal{L}(P_{1}P_{2}) &=&\omega
_{1}^{c}P_{1}(E_{3}^{c,0})P_{2}(E_{3}^{c,0})+\omega
_{2}^{c}P_{1}(E_{3}^{c,-2})P_{2}(E_{3}^{c,-2})+\omega
_{3}^{c}P_{1}(E_{3}^{c,2})P_{2}(E_{3}^{c,2})=0.  \notag
\end{eqnarray}%
Note that the last relation in (\ref{NN}) does not follow from the first
method.

As the final quantity we also compute the moment functionals defined in \cite%
{Favard,Finkel} as%
\begin{equation}
\mu _{n}:=\mathcal{L}(E^{n})=\sum\limits_{k=1}^{\ell }\omega
_{k}E_{k}^{n}=\sum\limits_{k=0}^{n-1}\nu _{k}^{(n)}\mu _{k},  \label{mu}
\end{equation}%
Once again also these quantities can be obtained in two alternative ways,
that is either from the computation of the integrals or directly from the
original polynomials $P_{n}$ and $Q_{n}$ without the knowledge of the
constants $\omega _{k}$. In the last equation the coefficients $\nu
_{k}^{(n)}$ are defined through the expansion $P_{n}(E)=2^{n-1}E^{n}-\sum%
\nolimits_{k=0}^{n-1}\nu _{k}^{(n)}E^{k}$ \ and $Q_{n}(E)=2^{n-1}E^{n-1}-%
\sum\nolimits_{k=0}^{n-2}\nu _{k}^{(n)}E^{k}$ for our even and odd
solutions, respectively. For the even solutions with $N=2+\lambda $ we obtain%
\begin{eqnarray}
\mu _{0}^{P} &=&1, \\
\mu _{1}^{P} &=&\lambda \zeta ^{2}, \\
\mu _{2}^{P} &=&\lambda ^{2}\zeta ^{4}-4\hat{\zeta}^{2}, \\
\mu _{3}^{P} &=&\lambda ^{3}\zeta ^{6}-12\lambda \zeta ^{2}\hat{\zeta}^{2}-16%
\hat{\zeta}^{2}, \\
\mu _{4}^{P} &=&\lambda ^{4}\zeta ^{8}-24\lambda ^{2}\zeta ^{4}\hat{\zeta}%
^{2}+16\left( \zeta ^{2}-1\right) ^{2}\zeta ^{4}-64\hat{\zeta}^{2},
\end{eqnarray}%
and similarly for the odd solutions with $N=3+2\lambda $ we compute for
instance%
\begin{eqnarray}
\mu _{0}^{Q} &=&1, \\
\mu _{1}^{Q} &=&4+\lambda \zeta ^{2}, \\
\mu _{2}^{Q} &=&16-4\hat{\zeta}^{2}+\lambda ^{2}\zeta ^{4}, \\
\mu _{3}^{Q} &=&\lambda ^{3}\zeta ^{6}-12\left( \lambda ^{3}+\lambda
^{2}+\lambda \right) \zeta ^{4}-48(2\lambda ^{2}+3\lambda +2)\zeta ^{2}+64.
\end{eqnarray}%
\qquad Thus $\mathcal{H}(N,\zeta ,\lambda )$ possesses indeed all the
standard features of a quasi-exactly solvable model of $E_{2}$-type.

\section{Conclusions}

Following the principles outlined in \cite{E2Fring} we have constructed a
new three-parameter quasi-exactly solvable model of $E_{2}$-type. One of the
parameters can be employed to interpolate between two previously constructed
models. With regard to one of the original motivations that triggered the
investigation of these models, that is the double scaling limit towards the
complex Mathieu equation, we found that for $\lambda =1$, i.e. $\mathcal{H}%
_{E_{2}}^{(1)}$, finite values for $N$ best approximate the complex Mathieu
system and mimic its qualitative behaviour. We provided a detailed
discussion of the determination of the exceptional points and the energy
branch cut structure responsible for the intricate energy loop structure
stretching over several Riemann sheets. The coefficient functions are shown
to possess the standard properties of weakly orthogonal polynomials.
\medskip 

\noindent \textbf{Acknowledgments:} I am grateful to Kazuki Kanki for making
reference \cite{thesisZhang} available to me.

\newif\ifabfull\abfulltrue

\end{document}